\documentclass[12pt]{iopart}

\usepackage{iopams}
\usepackage{graphicx}
\usepackage{amssymb}
\usepackage{graphicx}

\newcommand{\non}{\nonumber	\\}

\def\mbb{b\bar{b}}

\def\C{\chi_b}
\def\Cnp{\C(n{P)}}
\def\Cp{\C{(1P)}}
\def\Cpp{\C{(2P)}}
\def\Cppp{\C{(3P)}}
\def\U{\Upsilon}
\def\Uns{\U(n{S})}
\def\Us{\U{(1S)}}
\def\Uss{\U{(2S)}}
\def\Usss{\U{(3S)}}

\def\artanh{\mathop{\rm artanh}}
\def\bb{$\mbb$}
\def\c{$\C$}

\def\cnp{$\Cnp$}
\def\cp{$\Cp$}
\def\cpp{$\Cpp$}
\def\cppp{$\Cppp$}
\def\del{\partial}
\def\LambdaQCD{\Lambda_{QCD}}
\def\RAAQGP{R^{{QGP}}_{AA}}
\def\RAAQGPus{R^{{QGP}}_{AA}(\Us)}
\def\RAAuns{R_{AA}(\Uns)}
\def\RAAus{R_{AA}(\Us)}

\def\u{$\U$}
\def\uns{$\Uns$}
\def\us{$\Us$}
\def\uss{$\Uss$}
\def\usss{$\Usss$}

\begin{document}

\title{Bottomium suppression in PbPb collisions at LHC energies}

\author{Felix Nendzig and Georg Wolschin}

\address{Institut f{\"ur} Theoretische
Physik
der Universit{\"a}t Heidelberg, Philosophenweg 16, D-69120 Heidelberg, Germany, EU}
\ead{wolschin@uni-hd.de}
\begin{abstract}
We substantially refine our previously developed model for the suppression of $\Upsilon$ mesons in the quark-gluon plasma (QGP) formed in 2.76 TeV PbPb collisions at the LHC.
It accounts for gluodissociation of the six bottomium states \uns, \cnp\ with $n = 1,2,3$, collisional damping of these states as described in a complex potential, screening of the real part of the potential, and the subsequent decay cascade. In the hydrodynamical calculation of the expanding fireball, we take into account the effect of transverse expansion on the suppression factors, and finite transverse momenta of the heavy mesons in the medium. The running of the coupling is considered, resulting in larger gluodissociation decay widths. The initial central temperature is found to be 550 MeV at 0.1 fm/$c$. Our results are in good agreement with recent centrality-dependent CMS and ALICE data. The calculated suppression of the excited states relative to the ground state \us\ is not strong enough in peripheral collisions. Additional mechanisms are discussed. Predictions for 5.52 TeV are made.
\end{abstract}


\section{Introduction}	\label{sec:introduction}

The suppression of the yields of heavy quarkonia in relativistic heavy-ion collisions as compared to what is expected from $pp$ collisions at the same energy is a sensitive probe for the properties of the medium. The latter is likely to be a quark-gluon plasma (QGP) for sufficiently high incident energies (RHIC, LHC) and centralities that permit to attain an energy density above the critical one. Although most of the experimental and theoretical work in this field in the past 25 years was concerned with the $J/\Psi$ suppression and
its various possible origins, there are strong reasons to investigate bottomium suppression in the medium as a complementary process.

We had presented a corresponding theoretical investgation in \cite{nendzig-wolschin-2013}, where also the preference for the bottomium system was discussed: In particular the \us\ ground state with a mass of 9.46 GeV is very strongly bound, it melts as the last quarkonium state in the QGP only at about 4.10 $T_c $ \cite{wong-2005}. This turns it into a particularly interesting probe for processes like the dissociation by ultra soft thermal gluons in the QGP \cite{brezinski-wolschin-2012}, and collisional damping in the medium. The large $b$-quark mass renders theoretical derivations cleaner than in the charmonium system. And even at LHC energies, the number of $b\bar b$ pairs in the QGP remains small such that statistical recombination is unimportant, whereas it appears to be sizeable in the charmonium case.

As compared to our previous work \cite{nendzig-wolschin-2013},  a substantial number of refinements have been implemented into the model. As an example, we now use a different parametrisation of the relation between the number of collisions, number of participants, and the initial centrality-dependent temperature profile that employs the proportionality between the initial entropy density and the cube of the temperature, $s_0 \propto T_0^3$.

This modification produces more suppression and hence, changes the parameter space for the QGP lifetime, and Upsilon formation time that is consistent with the centrality-dependent CMS data for the ground state suppression. We also consider explicitly the finite transverse momentum of the bottomium in the QGP, with an ensuing blue shift in the direction of motion, and red shift in the opposite direction. In our description of the hydrodynamic expansion of the medium, we now explicitly include the effect of transverse expansion on the suppression factors, which leads to a slightly faster cooling of the system and hence, less suppression -- which is, however, compensated by the modified temperature profile.

Another major improvement concerns the consideration of the running coupling in the calculations of the wave functions, and the various dissociation processes. Since the coupling constant depends on the solution of the Schr\"odinger equation for the various bottomium states, we use an iterative method for the solution of the problem, together with the one-loop expression for the running of the coupling. In the calculation of the feed-down cascade that follows the suppression of the bottomium states in the medium, we now consider all six bottomium states \uns\ and \cnp\ with $n = 1$, 2, 3 below the $B\bar{B}$ threshold, including the recently discovered $\chi_b(3P)$ state as was proposed in \cite{vaccaro-etal-2013}.

In section~\ref{sec:dissociation} we employ a finite-temperature, complex potential model to calculate the bottomium wave functions and decay widths including gluodissociation in the QGP. In section~\ref{sec:hydrodynamics} a numerical, hydrodynamical calculation is presented to model the evolution of the temperature and velocity distributions of the fireball that includes the effect of transverse expansion. We then combine these results with the in-medium decay widths of chapter \ref{sec:dissociation} to calculate QGP-suppression factors $\RAAQGP$, that give the amount of suppression of the individual bottomium states within the fireball. After a decay cascade calculation has been performed, the final suppression factors $R_{AA}$ are presented in section~\ref{sec:comparison}, where they are compared to the most recent results of ALICE and CMS. We close with summary and conclusions in section~\ref{sec:conclusions}.

\section{Dissociation of bottomium}	\label{sec:dissociation}

In this section we calculate the in-medium color-singlet wave functions from an improved potential approach as compared to our earlier work \cite{brezinski-wolschin-2012,nendzig-wolschin-2013} that includes the effect of the running of the coupling with energy. The bound state wave functions $\psi_{nlm}(r,\theta,\varphi,T) = g_{nl}(r,T) Y_{lm}(\theta,\varphi)/r$, with the spherical harmonics $Y_{lm}$, are characterized by the principal, angular and magnetic quantum numbers $n$, $l$ and $m$, respectively. The \u\ and \c\ wave functions obey the temperature-dependent, radial, stationary Schr\"odinger equation
\begin{equation}
 \del_r^2 g_{nl}(r,T) =  m_b \bigg( V_{eff,nl}(r,T) - E_{nl}(T) + \frac{i\Gamma_{nl}(T)}{2} \bigg) g_{nl}(r,T)	\label{radialschroedinger}
\end{equation}
where $\Gamma_{nl}$ is the decay width, $E_{nl}$ the binding energy, $m_b$ the bottom mass and $V_{eff,nl}$ an effective interaction potential, that contains the centrifugal barrier and a complex interaction potential $V_{nl}$, whose real part vanishes at infinity due to screening,
\begin{equation}
 V_{eff,nl}(r,T) =  V_{nl}(r,T) + \frac{l(l+1)}{m_b r^2},
	\end{equation}

	\begin{equation}
 V_{nl}(r,T) = - \frac{\sigma}{m_D(T)} e^{-m_D(T)r}
	- C_F \alpha_{nl}(T) \Big( \frac{e^{-m_D(T)r}}{r} + iT \phi(m_D(T)r) \Big)
	\end{equation}
	\begin{equation}
		 \phi(x) = \int\limits_0^\infty \frac{dz \, 2z}{(1+z^2)^2} \left( 1 - \frac{\sin xz}{xz} \right),
 m_D(T) = T \sqrt{4\pi \alpha_s(2\pi T) \frac{2N_c + N_f}{6}},	\label{complex-potential}
\end{equation}
where the string tension equals $\sigma = 0.192$ GeV$^2$ and the Debye mass $m_D$ is obtained from perturbative HTL calculations. The complex potential (\ref{complex-potential}) is a combination of the potential found by \cite{laine-etal-2007,brambilla-etal-2008,beraudo-etal-2008} and the non-perturbative potential ansatz of \cite{karsch-etal-1988}.
The number of colors and flavors, respectively, are set to $N_c = N_f = 3$. The variable $\alpha_{nl}$ denotes the strong coupling $\alpha_s$ evaluated at the soft scale $S_{nl}(T)$,
\begin{equation}
 \alpha_{nl}(T) = \alpha_s(S_{nl}(T)), \quad S_{nl}(T) = \langle 1/r \rangle_{nl}(T).	\label{coupling}
\end{equation}
In this work we use the one-loop expression for the running of the coupling,
\begin{equation}
  \alpha_s(Q) = \frac{\alpha(\mu)}{1 + \alpha(\mu) b_0 \ln\frac{Q}{\mu}},	\quad	b_0 = \frac{11 N_c - 2 N_f}{6\pi},	\label{alpha_s}
\end{equation}
where $Q$ is the scale of four-momentum exchange and $\mu$ an arbitrary reference scale. Using values for $\alpha_s$ with matched charm- and bottom-masses \cite{bethke-2013} yields the QCD-scale $\LambdaQCD = 276.3$ MeV.

Eq. (\ref{radialschroedinger}) is solved numerically for the bottomium states \uns\ and \cnp\ ($n = 1$, 2, 3) and different temperatures. Since the coupling constant $\alpha_{nl}(T)$ depends on the solution $g_{nl}(r,T)$ of eq. (\ref{radialschroedinger}) we have to resort to an iterative procedure. First we choose a initial value $S^{(0)}$ for the soft scale in eq. (\ref{coupling}), specific to the state and temperature, and then evaluate eq. (\ref{radialschroedinger}) via a shooting method in the complex $(E,\Gamma)$-plane to obtain a first approximation of the wave function $g_{nl}^{(1)}$, energy $E_{nl}^{(1)}$ and decay width $\Gamma_{nl}^{(1)}$. We then recalculate the soft scale $S^{(1)}_{nl} = \langle 1/r \rangle_{nl}^{(1)}$ from $g_{nl}^{(1)}$ and use it together with $E_{nl}^{(1)}$ and $\Gamma_{nl}^{(1)}$ as initial values for the next step,
\begin{equation*}
 S^{(0)}	\rightarrow	g^{(1)},S^{(1)}, E^{(1)}, \Gamma^{(1)}	\rightarrow	\dots	\\
 \rightarrow	g^{(n)},S^{(n)}, E^{(n)}, \Gamma^{(n)}	\rightarrow	\dots
\end{equation*}
In this procedure, the bottom mass $m_b$ is fixed from the zero temperature case of the ground state, using the experimental value of the \us-mass \cite{PDG-2012} and replacing eq. (\ref{complex-potential}) by a Cornell potential. This yields

 $m_b = 4801$ MeV,
$ \alpha_{10}(0) = \alpha_s(1542 $ MeV) = 0.3984.

\noindent
Subsequently, for the other states and for finite temperature, $m_b$ is held fixed at this value and both $E$ and $\Gamma$ are varied in order to satisfy eq. (\ref{radialschroedinger}).

\begin{figure}[tp]
 \centering
  \includegraphics{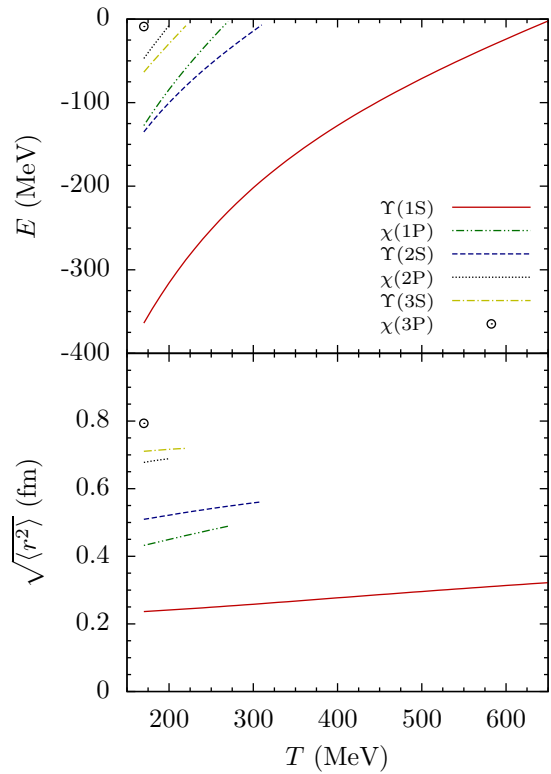}
\caption{Binding energies $E$ (top) and rms radii $\sqrt{\langle r^2 \rangle}$  (bottom) of the different bottomium states as a function of temperature.}
\label{fig:E-rms}
\end{figure}

Figure \ref{fig:E-rms} shows the binding energies $E_{nl}$ and the root-mean-square (rms) radii $\sqrt{\langle r^2 \rangle}$ for all six states under consideration. In the common picture color screening weakens the \bb-binding, causing the bound state to swell up to very large rms radii before it eventually dissolves. Remarkably the combined effect of color screening and collisional damping keeps $\sqrt{\langle r^2 \rangle}$ approximately constant even to the point of dissolution at the melting temperature $T_m$ (see table \ref{tab:melting-temperatures}).

\begin{table}[tp]
\caption{Melting temperatures $T_m$ of the different bottomium states. No bound state solutions to eq. (\ref{radialschroedinger}) exist for $T > T_m$.}
\label{tab:melting-temperatures}
 \vspace{3mm}
 \centering
\begin{tabular}{cr@{\hspace{2em}}}	\hline\hline
  State	&	\multicolumn{1}{c}{$T_m$ (MeV)}	\\	\hline
  \us   & 655	\\
  \cp   & 273	\\
  \uss   & 320	\\
  \cpp   & 206	\\
  \usss   & 228	\\
  \cppp   & $\sim 175$	\\	\hline\hline
\end{tabular}
\end{table}

\begin{figure}[t]
\centering
 \includegraphics{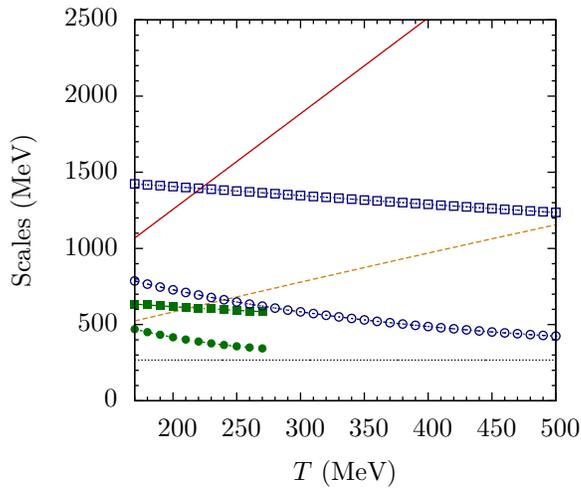}
\caption{Comparison of different scales in the potential model. The thermal scale $2\pi T$, the Debye mass $m_D$ and $\Lambda_{QCD}$ are plotted in solid, dashed and dotted lines, respectively. Soft scales $S_{nl}(T) = \langle 1/r \rangle_{nl}(T)$ (squares) and ultra soft scales $U_{nl} = \langle | V_{eff,nl} | \rangle(T)$ (circles) are plotted for the \us\ (empty) and \cp\ (filled) states.}
\label{fig:scales}
\end{figure}

In figure \ref{fig:scales} we have plotted different scales emerging in the potential model. In particular we compare the thermal scale $2\pi T$ with the soft scales $S_{nl}(T)$ of the \us\ and \cp. In the absence of the non-perturbative string part, the potential (\ref{complex-potential}) has been found to be valid only for $2\pi T \gg S_{nl}(T)$ in pNRQCD \cite{brambilla-etal-2008}. It is evident that this relation is indeed satisfied for the \cp, and likewise for all higher excited states, for all temperatures $T \geq T_c = 170$ MeV. For the \us, however, the potential (\ref{complex-potential}) should be replaced by a low-temperature expression for $T \lesssim 230$ MeV. Since our approach is guided by pNRQCD we should in principle also use two different potentials for the \us. However, we have found that the influence on the final results is negligible \cite{nendzig-phd} and so it is safe to use eq. (\ref{complex-potential}) only.

The ultra soft scale $U_{nl}$ is of the order of the binding energy $|E_{nl}|$ and hence, gluodissociation should be an ultra soft process since the gluon energy $E_g$ has to satisfy $E_g > |E_{nl}|$ in order to be able to dissociate the bottomium. The choice $U_{nl} = |E_{nl}|$, however, turns out to be unrealistic since $|E_{nl}| < \LambdaQCD$ for all excited states and even for the ground state \us\ if $T \gtrsim 240$ MeV \cite{nendzig-phd}. On the other hand, we find that $U_{nl} = \langle | V_{eff,nl} | \rangle$ represents a valid choice for the ultra soft scale, satisfying $U_{nl} > \LambdaQCD$ within a reasonable parameter regime as can be seen in figure \ref{fig:scales}.

\begin{figure}[t]
 \centering
  \includegraphics{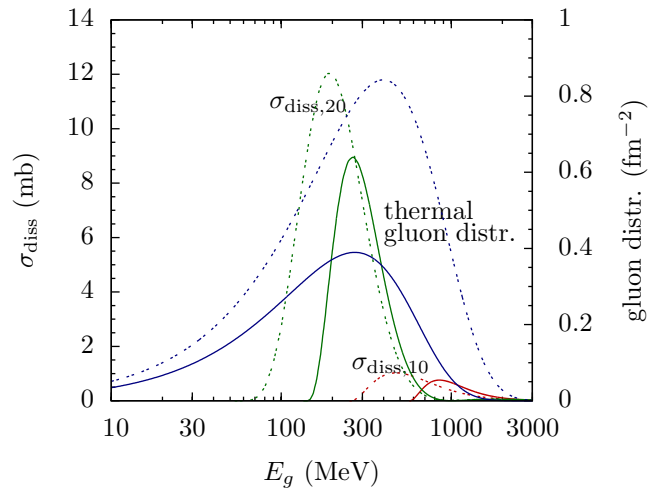}
\caption{Gluodissociation cross section $\sigma_{diss}$ (left scale) of the \us\ and \uss\ and the thermal gluon distribution (right scale) plotted for temperature $T=170$ (solid curves) and $250$ MeV (dotted curves) as functions of the gluon energy $E_g$.}
\label{fig:csec-thermal-overlap}
\end{figure}

The gluodissociation decay width is calculated by folding the singlet-octet dipole transition cross section $\sigma_{{diss},nl}$ with a Bose-Einstein distribution for the gluons \cite{brezinski-wolschin-2012,nendzig-wolschin-2013,brambilla-etal-2011},
\begin{equation}
 \Gamma_{{diss},nl}(T) \equiv \frac{g_d}{2\pi^2} \int\limits_0^\infty \frac{dE_g \, E_g^2 \, \sigma_{{diss},nl}(E_g)}{e^{E_g/T}-1},	\label{gamma-gd}
\end{equation}
where $g_d = 16$ is the number of gluon degrees of freedom. The width $\Gamma_{diss}$ depends on the overlap of the gluodissociation cross section with the gluon distribution. While the peak of the Bose-Einstein distribution moves to larger gluon energies with increasing temperature the opposite is the case for the shape of the cross section (see figure \ref{fig:csec-thermal-overlap}). In previous studies \cite{brezinski-wolschin-2012,nendzig-wolschin-2013}, where the running of the coupling was not considered in the Schr\"odinger equation, this behavior has resulted in a maximum of $\Gamma_{diss}$ at a certain temperature. This has led to the conclusion that gluodissociation would become inefficient at higher temperatures \cite{emerick-etal-2012,song-etal-2011,song-etal-2012}. In the present model, however, the decreasing ultra soft scale $U_{nl}$ enhances the coupling at higher temperatures and so results in a larger cross section at higher temperature.

\begin{figure}[t]
 \centering
  \includegraphics{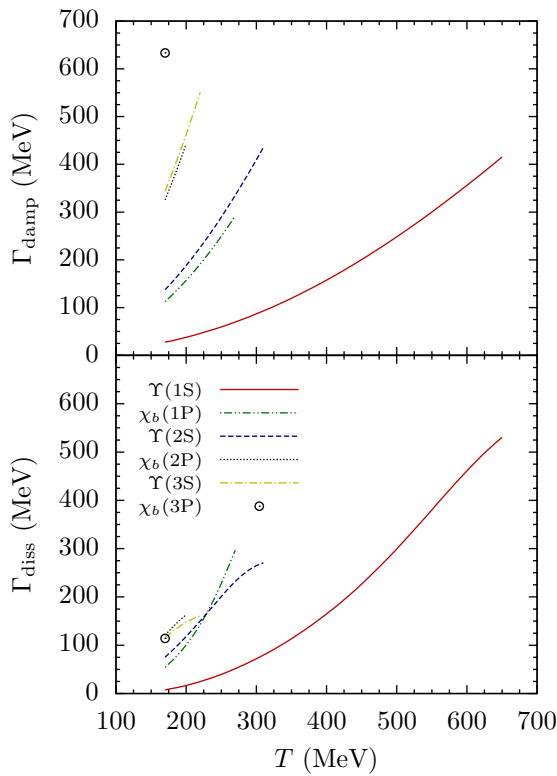}
\caption{Partial decay widths $\Gamma_{damp}$ as obtained from the Schr\"odinger equation (\ref{radialschroedinger}) and $\Gamma_{diss}$ due to gluodissociation are plotted versus temperature $T$ for the different bottomia.}
\label{fig:Gamma-part}
\end{figure}

The gluodissociation decay width therefore increases with temperature and does not achieve a maximum as shown in figure \ref{fig:Gamma-part}, where results for $\Gamma_{{diss},nl}$ are plotted together with the decay width $\Gamma_{nl} \equiv \Gamma_{{damp},nl}$, that originates from the imaginary part of the potential. It is evident that gluodissociation contributes significantly to the width at all temperatures. Only for the \cpp\ and \usss\ the shape of $\Gamma_{{diss}}$ appears to tend towards a maximum value as the melting temperature $T_m$ is approached.

\section{Hydrodynamical expansion}	\label{sec:hydrodynamics}

In order to calculate the amount of \u-suppression in the QGP using the melting temperatures and decay widths obtained in sec. \ref{sec:dissociation}, we need to model the evolution of the fireball produced in relativistic heavy ion collisions.

\begin{figure}[tp]
 \centering
  \includegraphics{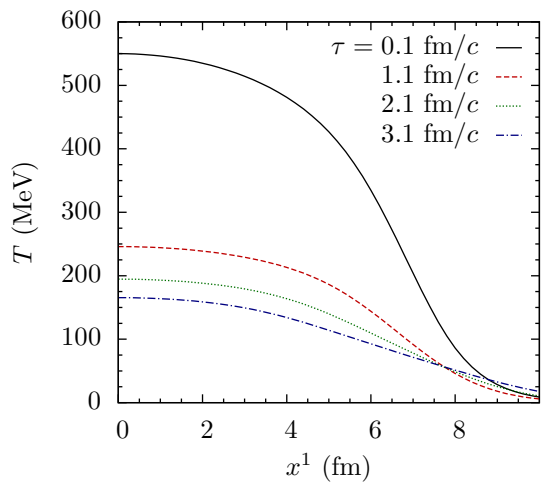}
\caption{Profiles of temperature $T$ in the fireball plotted along the $x^1$-axis for a central collision ($b = 0$) for times $\tau = 0.1$ (solid), 1.1 (dashed), 2.1 (dotted), 3.1 fm$/c$ (dash-dotted), respectively.}
\label{fig:profile-T}
\end{figure}

We describe the QGP by a relativistic, perfect fluid consisting of gluons and massless up-, down- and strange-quarks, whose energy-momentum tensor reads
\begin{equation}
 \mathcal{T} = (\varepsilon + P) u \otimes u + P,	\label{perfect-fluid-em-tensor}
\end{equation}
where $\varepsilon$ is the fluid's internal energy density, $P$ the pressure and $u$ the fluid four-velocity. For a general energy-momentum tensor the equations of motion are obtained by imposing four-momentum conservation, $\nabla \cdot \mathcal{T} = 0$, which yields
\begin{equation}
  \frac{1}{\sqrt{|\det g|}} \del_\mu \left( \sqrt{|\det g|} \mathcal{T}^\mu{}_\nu \right) =  \frac{1}{2} \mathcal{T}^{\mu\alpha} \del_\nu g_{\mu\alpha},	\label{energy-momentum-conservation}
\end{equation}
where $g = g_{\mu\nu} dx^\mu dx^\nu$ is the spacetime-metric and eq. (\ref{perfect-fluid-em-tensor}) has to be inserted for $\mathcal{T}$. The system of equations is closed by the equation of state, appropriate for a perfect, relativistic fluid,
\begin{equation}
 P = c_s^2 \varepsilon,		c_s = \frac{1}{\sqrt{3}},		\varepsilon = \varepsilon_0 T^4.	\label{EOS}
\end{equation}
We evaluate eq. (\ref{energy-momentum-conservation}) in the longitudinally co-moving frame (LCF), where the metric $g$ is given by
\begin{equation}
  g = -d\tau^2 + \tau^2 dy^2 + (dx^1)^2 + (dx^2)^2,	\label{metric}
\end{equation}
with the $x^1$-axis lying within and the $x^2$-axis orthogonal to the reaction-plane. In this frame the fluid flour-velocity $u$ reads
\begin{equation}
 u = \gamma_\perp (e_\tau + v^1 e_1 + v^2 e_2),	\non
 \gamma_\perp = \frac{1}{\sqrt{1 - (v^1)^2 - (v^2)^2}}.	\label{fluid-velocity}
\end{equation}
Note that the same transverse velocity components $v^1$, $v^2$ are measured in the laboratory frame (LF) as in the LCF; a property that is very convenient when dealing with quantities that depend on transverse momentum $p_T$. Inserting eqs. (\ref{perfect-fluid-em-tensor}) and (\ref{EOS}) - (\ref{fluid-velocity}) into eq. (\ref{energy-momentum-conservation}) yields
\begin{equation}
 \del_\mu (\tau T^4 u^\mu u_a) = - \frac{\tau}{4} \del_a T^4,	\non
 \del_\mu (\tau \, T^3 u^\mu) = 0,	\label{EOM}
\end{equation}
where the second equation corresponds to $u \cdot (\nabla \cdot \mathcal{T}) = 0$. We solve eqs. (\ref{EOM}) numerically, starting at the initial time $\tau_{init} = 0.1$ fm$/c$ in the LCF with the initial conditions
\begin{equation}
 v^1(\tau_{{init}}) = v^2(\tau_{{init}}) = 0
  \end{equation}
 \begin{equation}
 T(b, \tau_{{init}}, x^1, x^2) = T_0 \left( \frac{N_{mix}(b,x^1,x^2)}{N_{mix}(0,0,0)} \right)^{1/3}
 \label{T-initial-conditions}
 \end{equation}
 \begin{equation}
 N_{{mix}} = \frac{1 - f}{2} N_{{part}} + f N_{{coll}},
\end{equation}
where $f = 0.145$ \cite{PHOBOS-2004,ALICE-2011a}, $T_0 = 550$ MeV and the impact parameter $b$ (see \cite{nendzig-phd} for more information on the numerical procedure). Temperature profiles are shown in figure \ref{fig:profile-T} for a central collision ($b = 0$).

We define the QGP-suppression factor $R^{QGP}_{AA,nl}(c,p_T)$, which quantifies the amount of in-medium suppression of bottomia with transverse momentum $p_T$ for PbPb collisions in the centrality bin $c$, where $b_c \leq b < b_{c+1}$. The QGP-suppression factor is not directly measurable since it accounts only for the amount of suppression inside the fireball due to the three processes of color screening, collisional damping and gluodissociation. It is given by the ratio of the number of bottomia that have survived the fireball to the number of produced bottomia. The latter scales with the number of binary collisions at a given point in the transverse plane  and hence with the nuclear overlap, $N_{b\bar{b}} \propto N_{coll} \propto T_{AA}$. Thus we write $R^{QGP}_{AA}$ as follows:
\begin{equation}
 R_{AA,nl}^{QGP}(c,p_T) =	\non
 \hspace{-1cm}	\frac{\int_{b_c}^{b_{c+1}} db \, b \int d^2x \, T_{AA}(b,x^1,x^2) \, D_{nl}(b,p_T,x^1,x^2)}{\int_{b_c}^{b_{c+1}} db \, b \int d^2x \, T_{AA}(b,x^1,x^2)}.	\label{RAAQGP}
\end{equation}
The damping factor $D_{nl}$ is determined by the temporal integral over the total decay width $\Gamma_{{tot},nl} = \Gamma_{{damp},nl} + \Gamma_{{diss},nl}$,
\begin{equation}
 D_{nl}(b,p_T,x^1,x^2) =	\non
 \hspace{-1cm}	\exp\left[ - \int\limits_{\tau_{F,nl} \gamma_{T,nl}(p_T)}^\infty \frac{d\tau \, \Gamma_{{tot},nl}(T_{{eff},nl})}{\gamma_{T,nl}(p_T)} \right],	\label{damping-factor}
\end{equation}
where $\tau_{F,nl}$ is the formation time in the bottomium rest-frame, $\gamma_{T,nl}(p_T) = \sqrt{1 + (p_T/M^{{vac}}_{nl})^2}$ the Lorentz-factor due to transverse motion in the LCF, $M^{{vac}}_{nl}$ the experimentally measured bottomium vacuum mass and the effective temperature $T_{eff}$ will be properly defined below. In particular, for a bottomium state formed initially at the location $(x^1,x^2)$ that moves through the medium with the transverse velocity $\beta_{nl}(p_T)$, the exact functional form of the total decay width reads
\begin{equation}
  \Gamma_{{tot},nl} = \Gamma_{{tot},nl}( T_{{eff},nl}(b,p_T,\tau,x^1 + \beta^1_{nl} \tau,x^2 + \beta^2_{nl} \tau) ).
\end{equation}
Bottomium states are too massive to experience a substantial change of their momenta by collisions with the light medium particles. Hence there will be a finite relative velocity $v$ between the QGP and the \bb\ states. The relativistic Doppler shift will then result in an angle-dependent, effective temperature,
\begin{equation}
 T_{eff}'(v,\theta) = T \frac{\sqrt{1 - v^2}}{1 - v \cos\theta},	\label{Teff'}
\end{equation}
where $\theta$ is the angle between the direction of $\bf{v}$ and the scattering angle, as was also concluded in \cite[]{escobedo-etal-2011}.
In general $T_{{eff}}'$ results in a blue-shifted effective temperature in the forward direction and a red-shifted one in the backward direction. The effect of red and blue shift gets more and more pronounced with increasing velocity $v$ but the red-shifted region is growing, while the blue-shifted region is restricted to smaller and smaller angles $\theta$; a fact that has already been pointed out in \cite{escobedo-etal-2013}.

Here we approximate the effect of this anisotropic, effective temperature by averaging eq. (\ref{Teff'}) over the solid angle. In this way we obtain an isotropic, effective temperature $T_{eff}(v)$ as
\begin{equation}
 T_{eff}(v) = \frac{1}{4\pi} \int d\Omega \, T'_{eff}(v,\theta)
                            = T \, \sqrt{1 - v^2} \, \frac{\artanh v}{v} \leq T.
\end{equation}
We find that the effect of red shift dominates over blue shift for all relative velocities $v > 0$ (see \cite{nendzig-phd} for more information on the evolution of $T_{eff}$ in relativistic heavy ion collisions).

\begin{figure}[tp]
  \centering
	\includegraphics{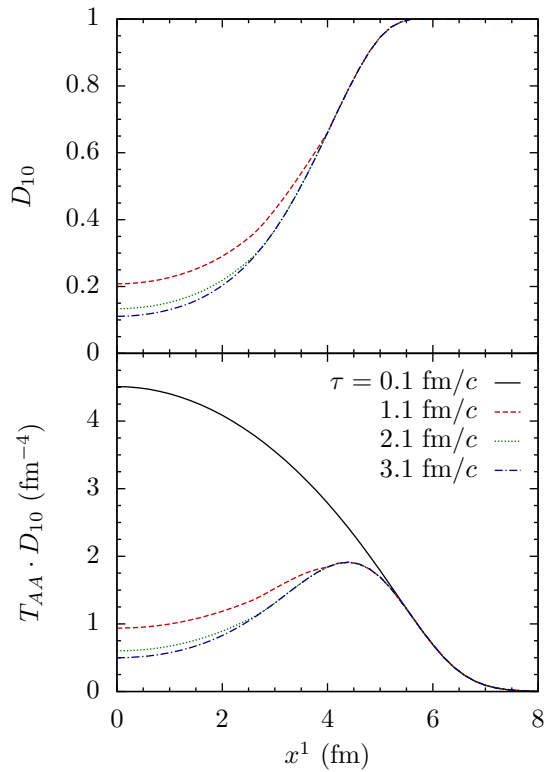}
\caption{Profiles of \us-damping factor $D_{10}$ (top) and $T_{AA} D_{10}$ (bottom) plotted along the $x^1$-axis for $p_T = 0$ in a central collision ($b = 0$) at times $\tau = 0.1$ (solid), 1.1 (dashed), 2.1 (dotted) and 3.1 fm$/c$ (dash-dotted).}
\label{fig:profile-D}
\end{figure}

Figure \ref{fig:profile-D} shows profiles of the \us-damping factor $D_{10}$ and the weighted damping factor $T_{AA} D_{10}$ that appears in the integrand of the numerator in eq. (\ref{RAAQGP}). The two quantities are plotted along the $x^1$-axis for a central collision ($b = 0$) for transverse momentum $p_T = 0$ and a formation time of $\tau_{10} = 0.5$ fm/$c$. In general bottomia dissolve very rapidly near the collision center, where the decay width is large. As the fireball cools, however, the dissociation rate decreases strongly.

\begin{figure}[tp]
\centering
\begin{picture}(250,300)
  \put(0,0){\framebox(248,140){\includegraphics[width=8.5cm]{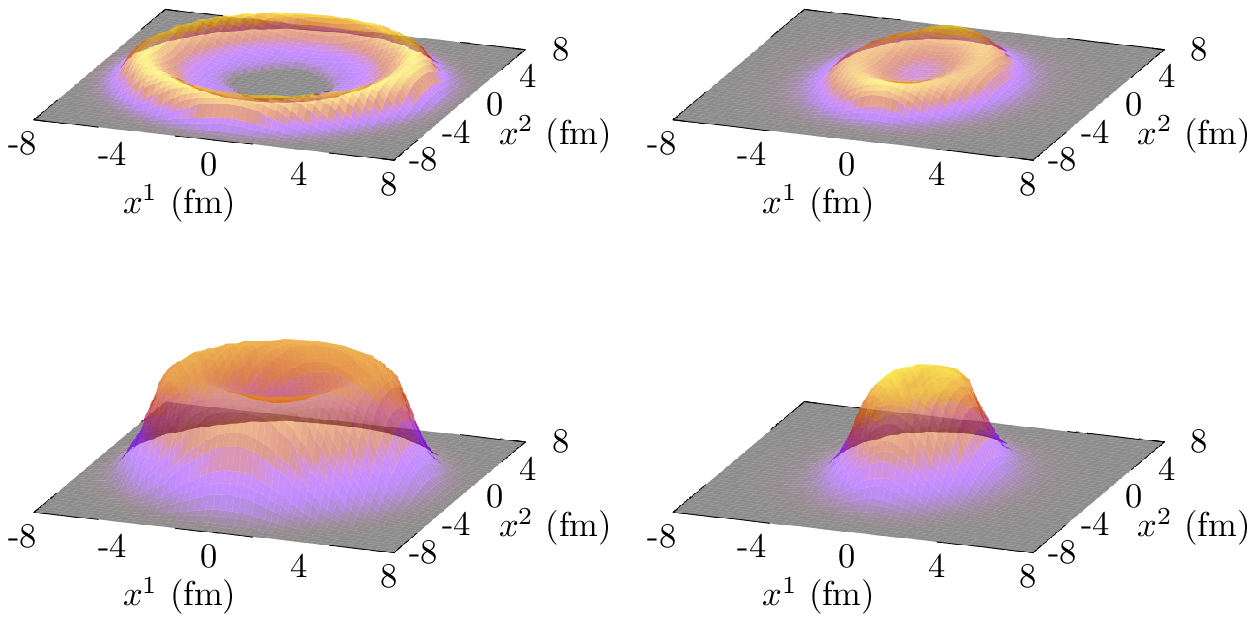}}}
  \put(0,145){\framebox(248,140){\includegraphics[width=8.5cm]{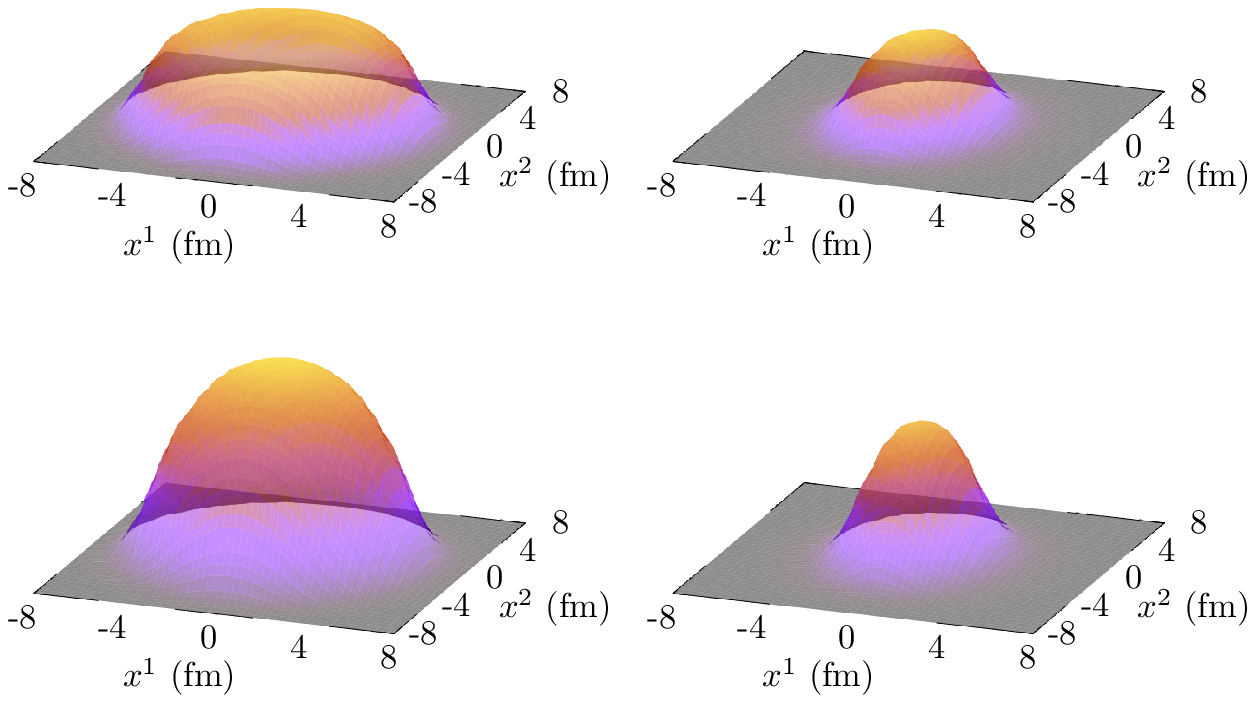}}}

  \put(40,290){\large \bfseries $\boldsymbol{b = 0}$ fm}
  \put(165,290){\large \bfseries $\boldsymbol{b = 8}$ fm}
  \put(110,65){\large \bfseries $\boldsymbol{\U}$(2S)}
  \put(110,210){\large \bfseries $\boldsymbol{\U}$(1S)}
\end{picture}
\vspace{0mm}
\caption{Two-dimensional profiles of $T_{AA} D_{n0}$ for the \us\ and \uss\ for central ($b = 0$, left-hand side) and peripheral collisions ($b = 8$ fm, right-hand side) with $p_T = 0$ and 12 GeV/$c$, respectively. From top to bottom row we have: (\us, $p_T = 0$), (\us, $p_T = 12$ GeV/$c$), (\uss, $p_T = 0$), (\uss, $p_T = 12$ GeV/$c$). The profiles scale with the fraction of \us\ and \uss\ that survive suppression throughout the lifetime of the QGP.}
\label{fig:profile-USupp}
\end{figure}

The weighted damping factor $T_{AA} D_{nl}$ directly scales with the number of surviving bottomia in the transverse plane. Hence it is very instructive to take a look at the two-dimensional plots of $T_{AA} D_{n0}$ shown in figure \ref{fig:profile-USupp}, where $\tau_{F,nl} = 0.5$ fm/$c$ has been used. The transverse $T_{AA} D_{n0}$-distributions of the \us\ and \uss\ are displayed for a central ($b = 0$) and a peripheral collision ($b = 8$ fm) for $p_T = 0$ and 12 GeV/$c$, respectively. Most bottomia are formed in the center of the heavy ion collision, where most binary nucleon-nucleon collisions occur. Due to the high central temperatures, however, strong suppression changes the shape of the surface from cone-like (peripheral) into volcano-like (central). Evidently, the \uss\ is suppressed much more efficiently than the more stable \us. Note that it is the action of color screening that forbids the formation of bound \bb\ states above the melting temperature $T_m$ and thus enforces $D_{nl} = 0$ in close proximity to the collision center. This does not occur for the \us\ since its melting temperature (see table \ref{tab:melting-temperatures}) is higher than the maximum fireball temperature, $T_m > T_0$. Large transverse momenta lead to significantly less suppression as time dilation causes bottomia to be formed at a later time in the LF/LCF where the QGP has already cooled down a bit in addition to the red-shifted effective temperature seen by the bottomia.

\begin{figure}[tp]
  \centering
 \includegraphics{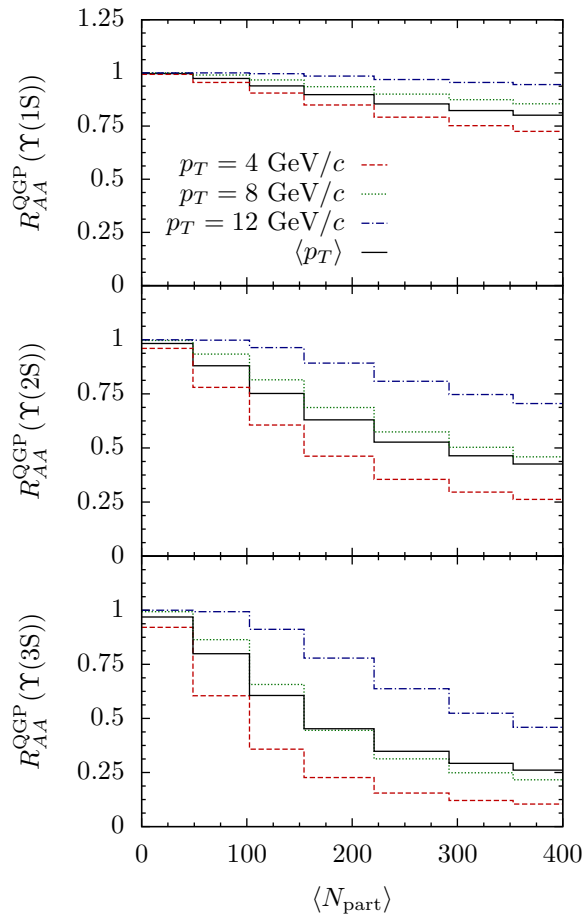}
\vspace{1mm}
\caption{QGP-suppression factors $\RAAQGP$ of the \us\ (top), \uss\ (middle) and \usss\ (bottom) as a function of centrality, for transverse momenta $p_T = 4$, 8, 12 GeV/$c$ (dashed, dotted and dash-dotted lines, respectively) and averaged over $p_T$ (solid lines).}
\label{fig:RAAQGP}
\end{figure}

Results for the QGP-suppression factors $R_{AA,n0}^{{QGP}} = R_{AA}^{{QGP}}(\Uns)$ are shown in figure \ref{fig:RAAQGP}, using $\tau_{F,nl} = 0.5$ fm/$c$, for transverse momenta $p_T = 4$, 8, 12 GeV/$c$ as well as $p_T$-averaged results. Lower $p_T$-values have been omitted for better comparison with the CMS data, which cover $p_T>4$ GeV/$c$.

The average over $p_T$ has been calculated from theoretical results for $p_T = 4$ up to 24 GeV/$c$ in steps of 2 GeV/$c$, according to
\begin{equation}
 \langle f \rangle_{nl} = \frac{\int_0^\infty dp_T \, \sigma_{nl}(p_T) f(p_T)}{\int_0^\infty dp_T \, \sigma_{nl}(p_T)},
\end{equation}
where we have used the bottomium production cross section $\sigma_{nl}(p_T)$ as measured by CMS \cite{CMS-2011a}, with $\sigma$(7-8 GeV/$c$) = 0.51 nb. The same cross section is assumed for the \c\ states as for the corresponding \u\ states.

Note that we have plotted $R_{AA}^{{QGP}}$ not as continuous function of centrality but averaged over centrality bins to achieve a better comparability with CMS data.

The ground state \us\ is significantly less suppressed than the \uss\ and \usss. The large in-medium decay widths of the excited states result in almost complete suppression in the QGP so that the overall QGP-suppression factor is non-zero only due to contributions from the peripheral collision regions where no QGP is formed. For high $p_T$, however, a significant fraction of the excited states is created in the boundary region of the QGP and can escape the plasma, seeing a strongly red-shifted, effective temperature.

\section{Comparison with LHC data}	\label{sec:comparison}

To obtain the final suppression factors that are to be compared with the data from the QGP-suppression factors we need to calculate the fractions of \uns\ that decay into dimuon pairs. Therefore we reconsider in this section the decay cascade of excited \bb\ states within the bottomium family. The suppression factor $R_{AA} = R_{AA}(\Uns,c,p_T)$, which compares the dimuon yields $N^{\mu}_n$ in PbPb- and $pp$-collisions for \uns\ states measured in the centrality bin $b_c \leq b < b_{c+1}$ with transverse momentum $p_T$, may be readily written as
\begin{equation}
 R_{AA}(\Uns,c,p_T) = \frac{ \sum_{I} \mathcal{M}_{nI} N^0_I R^{{QGP}}_{AA,I}(c,p_T) }{ \sum_{I} \mathcal{M}_{nI} N^0_I},
\end{equation}
where the matrix $\mathcal{M}$ describes the decay of \uns\ states into dimuon pairs, given the initial populations $N^0_I$ of the states $I = (n-l,l)$. In particular it is given by \cite{vaccaro-etal-2013,nendzig-phd}
\begin{equation}
  \mathcal{M}_{nI} = B_{\mu^\pm,n{S}} \mathcal{C}_{n{S},I},	\non
  \mathcal{C}_{IJ} = \left\{
\begin{array}{cc}
 \sum_{K = I + 1}^J B_{IK} \mathcal{C}_{KJ},	&	I < J	\\
 1,			&	I = J	\\
 0,			&	I > J
\end{array} \right.,
\end{equation}
where $B_{IJ}$ is the branching ratio for the decay $J \rightarrow I + X$ of the bottomium state $J$ into the lower lying state $I$ and $B_{\mu^\pm,n{S}}$ is the branching ratio for the decay $\Uns \rightarrow \mu^+ \mu^-$. While available experimental values are taken from the PDG \cite{PDG-2012}, the branching ratios of the \cppp\ have not been measured so we rely on theoretical predictions of the partial widths \cite{daghighian-silverman-1987}.

In order to obtain the initial populations we use CMS data from the LHC $pp$ run at $\sqrt{s_{NN}} = 2.76$ TeV where the relative yields of 1,  $0.56 \pm 0.13 { (stat)} \pm 0.02 { (sys)}$ and  $0.41 \pm 0.11 { (stat)} \pm 0.04 { (sys)}$ have been measured for dimuon pairs originating from \us-, \uss- and \usss-decays, respectively \cite{CMS-2012}. Further we make use of the CDF results for \us-feed down from \cnp\ states in $p\bar p$ collisions at $\sqrt{s} = 1.8$ TeV. It has been found that [$27.1 \pm 6.9$ (stat) $\pm 4.4$ (sys)]\% of the \us\ mesons come from \cp-decays, while [$10.5 \pm 4.4$ (stat) $\pm 1.4$ (sys)]\% come from \cpp-decays and the contribution from \cppp-decays is estimated to be less than 6\% \cite{CDF-2000}. In this work we will assume the \cppp-contribution to be at the estimated upper limit.

\begin{figure}[t]
 \centering
  \includegraphics{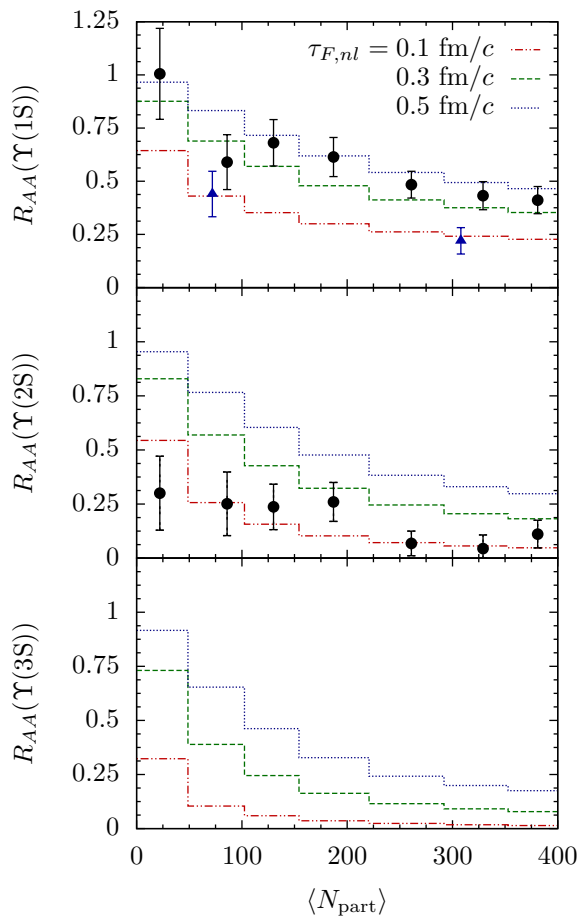}
\caption{Results for $R_{AA}$ of the \us\ (top), \uss\ (middle) and \usss\ (bottom) as a function of centrality, averaged over $p_T$, plotted for $\tau_{F,nl} = 0.1$, 0.3, 0.5 fm/$c$ (dash-dot-dotted, dashed, dotted lines, respectively), together with data from CMS \cite{CMS-2012}, circles, and ALICE \cite{castillo14}, triangles.}
\label{fig:RAA}
\end{figure}

The resulting suppression factors $\RAAuns$ are shown in figure \ref{fig:RAA} as functions of centrality. We have used $\tau_{{init}} = 0.1$ fm/$c$, $T_0 = 550$ MeV and performed the $p_T$-average from results for $p_T = 4$ up to 24 GeV/$c$ in steps of 2 GeV/$c$. Further we have used different formation times of $\tau_{nl} = 0.1$, 0.3, 0.5 fm/$c$. The theoretical results are compared to recent data of CMS \cite{CMS-2012} and ALICE \cite{manceau-2013,castillo14}. Note, however, that ALICE has measured at rapidity $2.5 < y < 4$ and transverse momentum $p_T > 0$ GeV/$c$, whereas CMS has measured at $|y|< 2.4$ and $p_T > 4$ GeV/$c$. There is slightly more suppression at the forward rapidities.
 Also, ALICE data are for the centrality bins 0 - 20\% and 20 - 90\%, respectively, as opposed to the CMS centrality bins, which we have used in this work to obtain the theoretical results. Minimum bias results are given in table \ref{tab:min-bias-results} for $\RAAuns$ and the double ratios $\frac{(\Uns/\Us)_{PbPb}}{(\Uns/\Us)_{pp}}$ together with CMS data \cite{CMS-2012}.

\begin{table*}[t]
 \centering
\caption{Minimum bias results for $\RAAuns$ and $\frac{(\Uns/\Us)_{PbPb}}{(\Uns/\Us)_{pp}}$.}
\label{tab:min-bias-results}
\begin{tabular}{c*{4}{l}}
\hline\hline
 $\tau_{F,nl}$ (fm$/c$) & 0.1 & 0.3 & 0.5 & CMS \cite{CMS-2012}	\\
\hline
 \vspace{-3mm}	\\
 $R_{AA}$(\us)   & 0.29 & 0.45 & 0.57 & $0.56 \pm 0.08 { (stat)} \pm 0.07 { (sys)}$	\vspace{1mm}	\\
 $R_{AA}$(\uss)  & 0.10 & 0.29 & 0.43 & $0.12 \pm 0.04 { (stat)} \pm 0.02 { (sys)}$	\vspace{1mm}	\\
 $R_{AA}$(\usss) & 0.04 & 0.16 & 0.29 & $0.03 \pm 0.04 { (stat)} \pm 0.01 { (sys)}$	\vspace{1mm}	\\
 $\frac{(\Uss/\Us)_{PbPb}}{(\Uss/\Us)_{pp}}$ & 0.34 & 0.64 & 0.74 & $0.21 \pm 0.07 { (stat)} \pm 0.02 { (sys)}$	\vspace{1mm}	\\
 $\frac{(\Usss/\Us)_{PbPb}}{(\Usss/\Us)_{pp}}$ & 0.13 & 0.33 & 0.49 & $0.06 \pm 0.06 { (stat)} \pm 0.06 { (sys)}$	\vspace{1mm} \\
\hline\hline
\end{tabular}
\end{table*}

\begin{figure}[t]
 \centering
  \includegraphics{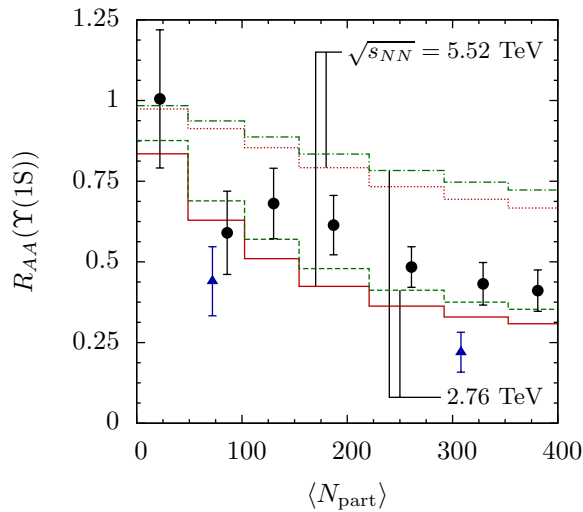}
\caption{Predictions for $\RAAus$ and $\RAAQGPus$ for PbPb collisions at $\sqrt{s_{NN}} = 5.52$ TeV (solid and dotted lines, respectively) and the previous results for $\sqrt{s_{NN}} = 2.76$ TeV (dashed and dash-dotted lines, respectively) as a function of centrality, averaged over $p_T$, plotted for $\tau_{F,nl} = 0.3$ fm/$c$, together with data from CMS \cite{CMS-2012}, circles, and ALICE \cite{castillo14}, triangles, for $\sqrt{s_{NN}} = 2.76$ TeV.}
\label{fig:RAA-552}
\end{figure}

In figure \ref{fig:RAA-552} we display predictions for $\RAAus$ and $\RAAQGPus$ for PbPb collisions at $\sqrt{s_{NN}} = 5.52$ TeV. We have used the scaling relation between the initial entropy density and the charged particle multiplicity per unit rapidity, $s_0 \propto dN_{ch}/d\eta$ \cite{bjorken-1983,baym-etal-1983,gyulassi-matsui-1984}, and inserted $s_0 \propto T_0^3$ together with extrapolated results for $dN_{ch}/d\eta$ \cite{wolschin-2013} to obtain an increase by 6.6\% in the initial temperature, i.e. $T_0 = 586$ MeV in our case. We have used $\tau_{nl}$ = 0.3 fm/$c$ and left the other parameters at the same values as for figure \ref{fig:RAA}.

The \us\ ground state is found to be slightly more suppressed in the medium at the higher energy of 5.52 TeV and hence, also the total suppression of the \us\ in PbPb as compared to $pp$ is stronger, but this enhancement is less than 10 \%: Doubling the energy in PbPb collisions at LHC is not expected to have a dramatic effect on bottomium suppression in heavy-ion colliisions.

\section{Summary and conclusions}	\label{sec:conclusions}

Our model for bottomium suppression in the quark-gluon plasma that had been outlined in \cite{nendzig-wolschin-2013} is substantially refined in the present work. The general approach is based on the realisation that, due to the strong binding of bottomia, and in particular the \us  state, only very few processes contribute significantly to their suppression in relativistic heavy ion collisions. Therefore we consider only the three processes of color screening, gluodissociation and collisional damping -- plus the substantial effects of the subsequent feed-down cascade. The treatment of the latter is performed as detailed in \cite{vaccaro-etal-2013} -- with the inclusion of the recently detected $\chi_b(3P)$ state, but without further modifications. A proper treatment of the cascade shows that it causes a large fraction of the suppression for all centralities because in the quark-gluon plasma the excited states are mostly suppressed and hence, the normal feed-down to the lower-lying states is missing.

As a major improvement of our previous work \cite{brezinski-wolschin-2012,nendzig-wolschin-2013} we now include the running coupling when solving the Schr\"odinger equation to obtain the bottomium wave functions, and the dissociation cross sections. Not only does the running of the coupling increase the melting temperatures, especially for the excited states, but the increased ultra soft coupling also results in a significantly larger gluodissociation decay width, and a monotonic rise with temperature in the physically relevant region. (Without the running coupling, the gluodissociation width acquires a maximum as a function of temperature).

We have also considered the finite transverse momenta of bottomia and their relative velocities with respect to the expanding QGP that surrounds the quarkonia. The ensuing relativistic Doppler shift results in an angle-dependent effective temperature, causing a blue shift in the forward and a red shift in the backward direction of the moving bottomium, which is considered in the calculation of the bottomium suppression in the QGP. When averaging over the solid angle, the red shift is actually found to be more important than the blue shift for all finite relative velocities, leading to a lower temperature profile and hence, less suppression. This is, however, more than compensated by a more realistic initial temperature distribution that makes use of the proportionality between the initial entropy density and the cube of the temperature. It depends on the third root of a mixture of the number of participants and the number of collisions. It is supported by experimental results from PHOBOS \cite{PHOBOS-2004} and ALICE \cite{ALICE-2011a}.

The hydrodynamical model of the fireball is refined to account for transverse expansion, which leads to a faster cooling of the system. Again, this is compensated by the now more realistic initial conditions, eq.(\ref{T-initial-conditions}). In view of the in many instances good agreement of ideal hydrodynamics with bulk observables in relativistic heavy-ion collisions, we conjecture that its use for the description of bottomium suppression in the medium is permissible, provided both longitudinal and transverse expansion are properly considered.

Due to the compensating effects in the above refinements of our model as compared to \cite{nendzig-wolschin-2013}, it turns out that the parameter space that is compatible with the
centrality-dependent \us\  suppression data as measured by CMS \cite{CMS-2012} is indeed close to the values that we had obtained previously, where we had used a bottomium formation time of 0.1 fm/$c$ and QGP lifetimes of 4-8 fm/$c$. However, in this work the improved initial conditions cause a flatter temperature profile with substantially more suppression in the QGP, which is counteracted by larger formation times of the order of 0.4 fm/$c$ (see figure 9): This circumvents the initially extremely hot (but rapidly cooling) zone and hence, reduces the suppression such that it agrees with the \us\ data. Instead of the QGP lifetime we now use the initial central temperature $T_0$ as a parameter, see eq. (\ref{T-initial-conditions}). We find $T_0 = 550$ MeV for 2.76 TeV PbPb, and obtain an excellent description of the centrality-dependent \us\ suppression data \cite{CMS-2012}.

When implementing the above improvements into our model, we had also intended to obtain a better theoretical understanding of the suppression of the \uss\ state as a function of centrality:
In \cite{nendzig-wolschin-2013} we had found that in peripheral collisions much less suppression is predicted than was measured by CMS \cite{CMS-2012}. As is shown in figure 9, there is indeed a somewhat larger suppression in the calculation for the \uss\ state, but a simultaneous precise modeling of the centrality dependence for both states is difficult to achieve by changing the model parameters formation time and/or initial temperature. A refined treatment using state-dependent formation times may resolve the issue, since shorter formation times for the excited states cause stronger suppression. However, this would also result in additional suppression of the \us\ state by missing feed-down. Consequently the suppression pattern can only be improved by means of modified formation times if a good balance between these two effects can be found. In addition, such an approach should be built upon a theory for state-dependent formation times, which has not yet been developed.

As a more promising answer to the conundrum, one could resort to additional suppression mechanisms which we have not yet accounted for, and which act differently on the ground and excited states. Cold nuclear matter (CNM) effects may indeed provide the solution, but nuclear shadowing is expected to act on ground and excited states in a similar fashion.
A remaining possibility is hadronic dissociation -- mostly by the large number of pions in the final state even in more peripheral collisions --, which is likely to be relevant only for the excited bottomium states, but not for the strongly bound spin-triplet ground state.  In any case, the current picture has to be refined if higher accuracy is desired. The broad range of different approaches \cite{emerick-etal-2012,song-etal-2012,strickland-2011,strickland-bazow-2012,alford-strickland-2013} in the field shows that a full account of the phenomenon of \u\ suppression in relativistic heavy ion collisions at the LHC remains a very ambitious goal.

Given the reliable modelling of the \us\  suppression in 2.76 TeV PbPb, we have also made a prediction at the LHC design energy of 5.52 TeV, where the initial central temperature is increased by 6.6\% from 550 MeV to 586 MeV. This leads to a slightly stronger suppression, but the effect is less than 10\% at all centralities. Whereas the  investigation of bottomium suppression at the highest LHC energy will thus be interesting regarding the energy dependence of the balance between ground and excited states, one should not expect dramatic modifications of the ground-state suppression.



\vspace{.4cm}
\bf{Acknowledgments}\\
\rm

This work has been partially supported by the IMPRS-PTFS and the ExtreMe Matter Institute EMMI.
F.N. is now at Funkinform, Ettlingen, Germany.


\section*{References}
\bibliographystyle{jphysg}
\bibliography{references}

\end{document}